\documentclass[twocolumn,trackchanges]{aastex62}
\usepackage{hyperref}
\hypersetup{bookmarksnumbered=true, bookmarksopen=true}
\usepackage{amssymb,amsmath,amsfonts}
\usepackage{footnotebackref}

\allowdisplaybreaks[1]

\usepackage{savesym}
\savesymbol{tablenum}
\usepackage{siunitx}
\restoresymbol{SIX}{tablenum}
\usepackage{chngcntr} %to reset appendix figure counters

\usepackage{bm}
\usepackage{booktabs}
\usepackage{paralist}

\usepackage{newtxtext}
\usepackage[varvw]{newtxmath}

\graphicspath{{figs2/}}

\newcommand{\refsec}[1]{Section~\ref{#1}}
\newcommand{\reffig}[1]{Fig.~\ref{#1}}

\newcommand{\refeqn}[1]{Equation~(\ref{#1})}

\renewcommand{\vr}{v_{r}}

\newcommand{\rhom}{\rho_{\mathrm{m}}}
\newcommand{\rd}{r_{\mathrm{id}}}

\newcommand{\rta}{r_{\mathrm{ta}}}

\newcommand{\env}{\eta_{\mathrm{env}}}

\newcommand{\M}[1]{M_\mathrm{#1}}    % e.g. \M{200c}
\newcommand{\R}[1]{R_\mathrm{#1}}    % e.g. \M{200c}
\newcommand{\V}[1]{V_\mathrm{#1}}    % e.g. \M{200c}

\newcommand{\msun}{M_{\odot}}

\newcommand{\mpc}{\mathrm{Mpc}}
\newcommand{\kpc}{\mathrm{kpc}}
\newcommand{\kms}{\mathrm{km \, s}^{-1}}
\newcommand{\msunh}{h^{-1} M_{\odot}}
\newcommand{\mpch}{h^{-1} \mathrm{Mpc}}

\shorttitle{Milky Way boundary}
\shortauthors{Li and Han}
\begin{document}
%% =====================================

\title{\large\textbf{The outermost edges of the Milky Way halo from galaxy kinematics}}

\author[0000-0001-7890-4964]{Zhao-Zhou Li}
\altaffiliation{lizz.astro@gmail.com}
\affil{Department of Astronomy, School of Physics and Astronomy, Shanghai Jiao Tong University, Shanghai 200240, China}
\affil{Shanghai Key Laboratory for Particle Physics and Cosmology, Shanghai 200240, China}

\author[0000-0002-8010-6715]{Jiaxin Han}
\altaffiliation{jiaxin.han@sjtu.edu.cn}
\affil{Department of Astronomy, School of Physics and Astronomy, Shanghai Jiao Tong University, Shanghai 200240, China}
\affil{Shanghai Key Laboratory for Particle Physics and Cosmology, Shanghai 200240, China}

%% =====================================
\begin{abstract}
We measure for the first time the outermost edges of the Milky Way (MW) halo in terms of the depletion and turnaround radii. The inner depletion radius, $r_\mathrm{id}$, identified at the location of maximum infall velocity, separates a growing halo from the draining environment, while the turnaround radius, $r_\mathrm{ta}$, marks the outermost edge of infalling material towards the halo, both of which are located well outside the virial radius. Using the motions of nearby dwarf galaxies within $3\mathrm{Mpc}$, we obtain a marginal detection of the infall zone around the MW with a maximum velocity of $v_\mathrm{inf, max}=-46_{-39}^{+24}\mathrm{km s^{-1}}$. This enables us to measure $r_\mathrm{id}=559\pm 107 \mathrm{kpc}$ and $r_\mathrm{ta}=839\pm 121 \mathrm{kpc}$. The measured depletion radius is about 1.5 times the MW virial radius ($R_\mathrm{200m}$) measured from internal dynamics.
Compared with halos in the cosmological simulation Illustris TNG100,
the factor 1.5 is consistent with that of halos with similar masses and dynamical environments to the MW but slightly smaller than typical values of Local Group analogs, potentially indicating the unique evolution history of the MW.
These measurements of halo edges directly quantify the ongoing evolution of the MW outer halo and provide constraints on the current dynamical state of the MW that are independent from internal dynamics.
\end{abstract}


% =====================================
\section{Introduction}

The Milky Way (MW) galaxy is one of the most important laboratories for studying galaxy formation and cosmology,
given the abundant information available from its well-resolved constituents \citep{Bland-Hawthorn2016a}.
In the current hierarchical structure formation framework,
the properties of a galaxy are tightly connected to the properties of its dark matter halo.
To place the MW in the context of cosmological galaxy formation,
one usually relies on the estimated size of the MW halo according to a certain definition of the halo boundary and the corresponding enclosed mass.

Despite many efforts dedicated to measuring the mass distribution 
in the virialized region of the MW halo \citep{Wang2019b} in observations, 
much less attention has been paid to the very outskirts beyond the formal virial radius.
In addition to the normally higher incompleteness and larger measurement errors for tracers at large distances,
the lack of equilibrium in this region also blocks dynamical modeling attempts based on the steady-state assumption~\citep{oPDFI,oPDFII} and thus requires better theoretical understanding.

Conventionally, most studies use the classical
virial definition (or its variants) derived from the spherical collapse model \citep{Gunn1972},
which marks out a radius by a fixed enclosed overdensity under some idealized assumptions.
However, a halo in the real universe is not abruptly separated from the neighboring environment at this specific radius.
In fact, the mass distribution within and around a halo is a continuous mixture of
the virialized content, the infalling materials, and background materials receding with the rest of the universe.
This fact has inspired people to further investigate other boundaries better separating these components
(see \citealt{Fong2020} for a detailed summary),
such as the splashback radius \citep{Adhikari2014,Diemer2014,Diemer2017,Aung2021}, 
the depletion radius \citep{Fong2020},
and the turnaround radius \citep[e.g.,][]{Gunn1972,Cuesta2008,Pavlidou2014,Faraoni2015},
from the inside out.
Unlike the spherical overdensity-based definition,
the latter boundaries are more directly associated with dynamical processes,
and hence detectable from the kinematics of tracers \citep[e.g.,][]{Deason2020,Bose2020,Tomooka2020}. 
This is a particular advantage because 
we can measure the velocity of tracers, e.g., nearby galaxies, even at a large distance,
but cannot observe the density directly.

The different halo radii definitions also serve to provide different insights on the structure and evolution of halos. In a recent work, \citet{Fong2020} introduced the \textit{inner depletion radius}, $\rd$, defined at the location of the maximum mass inflow rate, as the outer edge of the \emph{growing} part of a halo. 
Practically, this radius is identifiable at the location of the maximum infall velocity (see Fig.~11 of \citealt{Fong2020}) which is the approach we follow in this work.
With $\rd$ defined at the maximum inflow location, matter within $\rd$ gets deposited onto the halo as the infall rate slows down towards the inner halo.
Outside this radius, however, matter is being pumped into the halo and gradually depleted due to the increasing infall rate towards the inner region. 
This process leads to the formation of a relatively flat shoulder in the density profile and a trough in the bias profile around the $\rd$ scale (\citealt{Fong2020}).
Thus, this location marks the transition between the halo being built up and the environment being depleted by halo accretion.
Moreover, the enclosed density within this radius is found to have an approximately universal value, which enables us to easily estimate the enclosed mass.

From the perspective of particle orbits, $\rd$ can be interpreted as a boundary enclosing a more complete population of splashback orbits than the customary \textit{splashback radius} defined at the steepest slope radius, $r_{\rm sp}$. The latter is based on the steepening in the slope resulted from the buildup of particles at their first orbital apogees, but it is found to enclose only about 75\% of the splashback orbits \citep{Diemer2017}.
Hence, $\rd$ is normally outside the splashback radius, $r_{\rm sp}$, with $\rd\approx 1.7 \sim 2.6 r_{\rm sp}$.\footnote{This relation is converted combining the relations of $\rd\approx 0.85 r_{\rm cd}$ and $r_{\rm cd}=2-3 r_{\rm sp}$ in \citet{Fong2020}, where $r_{\rm cd}$ is the characteristic depletion radius defined at the minimum bias.}
Interestingly, this scale is shown to be very close to (or $\sim\!\! 15$ percent smaller than) the location of the minimum in the halo bias profile \citep{Han2018} around the trans-linear scale and  almost identical to the optimal halo exclusion radius measured by \citet{Garcia2020} that defines the geometrical boundary of non-overlapping halos in the halo model description of the large-scale structure.

Compared with the virial radius, $\rd$ is roughly located at the $1.6 \R{200m}$, where the $\R{200m}$ is the radius within which the average density is 200 times the mean background density.%
\footnote{
Similarly, $\R{200c}$ and $\R{vir}$ are defined as the radius within which the average density is 200 and $\Delta_\mathrm{vir}$ times the critical density of the universe, respectively, where $\Delta_\mathrm{vir}$ is the virial overdensity predicted from the spherical collapse model \citep{Bryan1998}.
}
By definition, the inner depletion radius at maximum infall is enclosed within the turnaround radius where the radial velocity reaches zero. The turnaround radius is of important dynamical significance as it separates infalling material from the expansion of the universe, and can serve as a probe of both halo evolution and the background cosmology~\citep[e.g.,][]{Gunn1972,Cuesta2008,Pavlidou2014,Faraoni2015}.

In this work, we present the first measurement of the inner depletion radius of the MW
using the motion of nearby dwarf galaxies, along with the turnaround radius measured from the same data set.
Although these radii were first introduced based on dark matter, galaxies are found to closely trace the underlying phase space structures of dark matter \citep[e.g.,][]{Han2020,Deason2020} especially in the outskirts of haloes. As a result, we will use galaxies as tracers to probe these radii. The measurements are then compared directly with those using galaxies in hydrodynamical simulations, as well as with previous results using dark matter particles.
Using the scaling relation learned from halos in simulations,
the enclosed masses within these boundaries are also estimated. As these boundaries directly quantify the ongoing evolution of the MW halo, the measurements can provide crucial information for better placing the MW into a cosmological context of halo evolution and galaxy formation.

The structure of this letter is as follows.
We present the measurements of the MW's outer edges in \refsec{sec:mw}, 
interpret the results with simulations in \refsec{sec:validate},
compare them with previous measurements in \refsec{sec:compare},
and summarize in \refsec{sec:conclusion}.
In addition, we provide the details of measuring the velocity profile in Appendix \ref{sec:gp}
and selecting simulation sample in Appendix \ref{sec:simu}.

%% =====================================
\section{The outer edges of the MW}\label{sec:mw}

We use nearby galaxies within $3\mpc$ of the MW, compiled from the catalog of the Local Volume galaxies 
\citep{Karachentsev2013,Karachentsev2019}%
\footnote{\url{http://www.sao.ru/lv/lvgdb/tables.php}, updated on 2020-08-12}
and the catalog of Nearby Dwarf Galaxies \citep{McConnachie2012}%
\footnote{\url{http://www.astro.uvic.ca/\~alan/Nearby\_Dwarf\_Database.html}, updated on 2021-01-19}.
The observed Heliocentric line-of-sight velocities are converted into radial velocities in the Galactocentric rest frame.
The proper motions from the catalog of Nearby Dwarf Galaxies 
(mostly measured by \citealt{McConnachie2020}) are used for the conversion when available.
For the remaining galaxies, we ignore their proper motions in the conversion considering their large distance.
The observational error of the line-of-sight velocity is typically smaller than several $\kms$,
which is negligible in this task compared with the bulk motion at several tens or hundreds of $\kms$ level.

\begin{figure}[bt]
\centering
\includegraphics[width=0.47\textwidth]{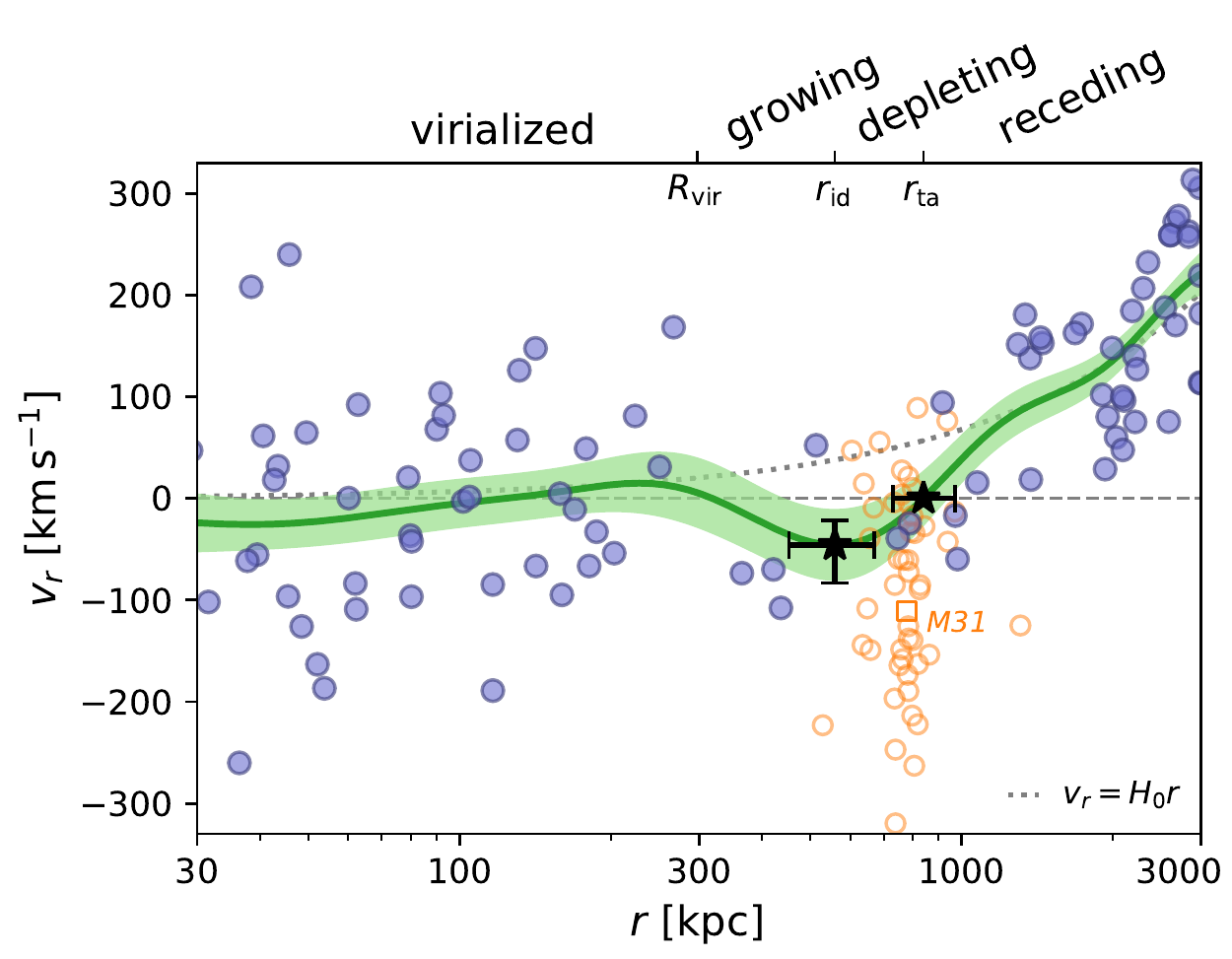}
\includegraphics[width=0.47\textwidth]{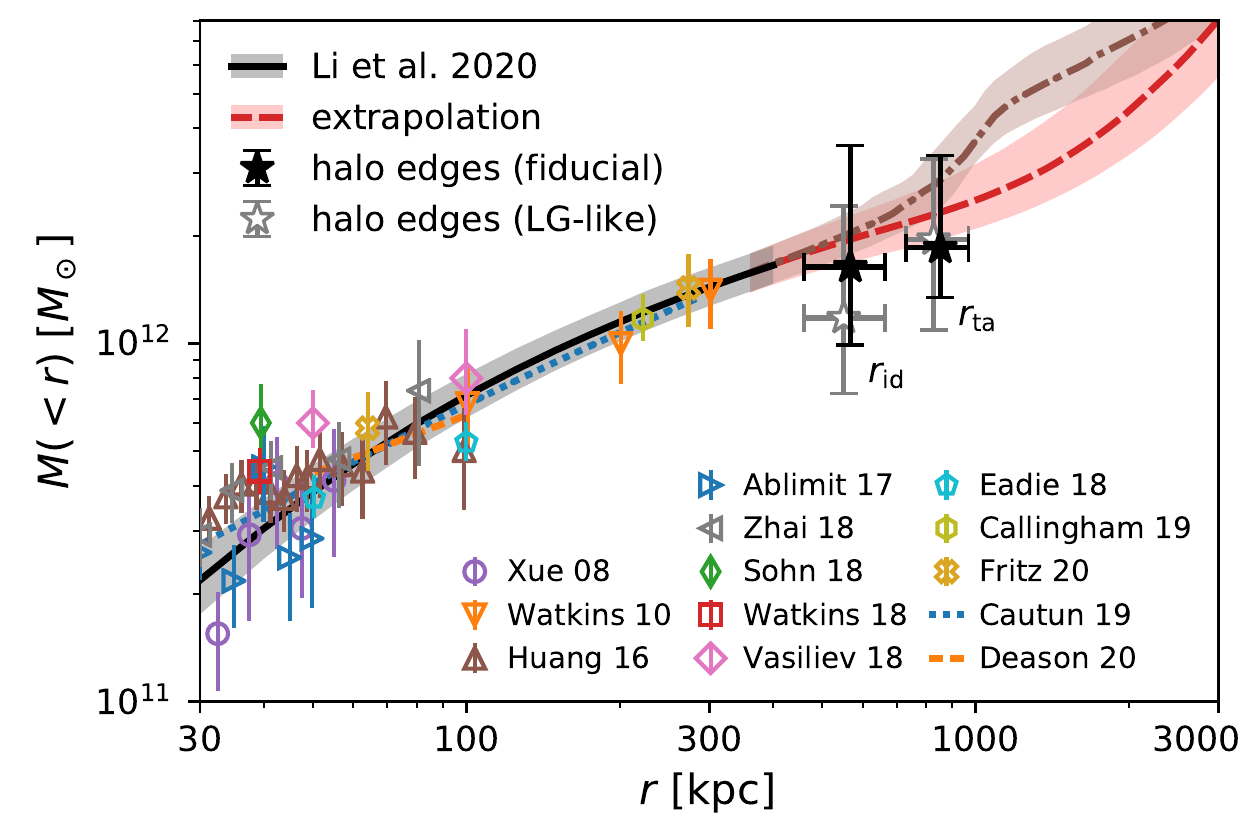}
\caption{%
    Top panel:
    Radial velocities of galaxies within 3 Mpc of the MW.
    Galaxies within $600\kpc$ from M31 are marked as
    open circles and discarded in the analysis.
    The mean velocity profile (green solid curve) and its $1\sigma$ uncertainty (green band) are computed from the remaining galaxies (filled circles).
    The measured MW edges including
    the \textit{inner depletion radius} (i.e., location of maximum infall), $r_\mathrm{id}$, 
    and the \textit{turnaround radius}, $r_\mathrm{ta}$, are indicated by star symbols.
    The Hubble flow, $\vr=H_0 r$, is shown by the dotted line for reference.
    Bottom panel:
    The MW mass profile.
    The star symbols indicate the estimated MW masses within the corresponding edges based on their typical enclosed densities in simulation. The estimates calibrated using a fiducial sample (black) or LG-like sample (gray, slightly shifted horizontally for clarity) of halos in simulation are shown separately (see the text for detail). Previous measurements of the inner MW mass profile using stars, globular clusters, and satellite galaxies 
    are shown for comparison. As an extrapolation of the inner profile \citep{Li2020},  the long-dashed (dash-dotted) curve shows the mean mass profile of the fiducial (LG-like) halos in the TNG100 simulation.
    The error bars or shades correspond to the 68\% confidence intervals.
}
\label{fig:local_group}
\end{figure}

The Galactocentric distances and radial velocities, $\{r, \vr\}$, of these galaxies are shown in
\reffig{fig:local_group}. 
In this work, we exclude galaxies within 600 $\kpc$ from M31 (about $1.5 \R{200m, M31}$) to reduce the potential influence of our massive neighbor. 
We have also checked that our results are not very sensitive to the particular choice of this radius of exclusion.
Changing the exclusion radius from 550 to 850 kpc only leads to a variation $\lesssim 2\%$ in the measured edges, while using a smaller value (e.g., 400 kpc) leads to slightly larger estimates (by factors of 5\% in $\rd$ and 10\% in $\rta$). Note the six dwarf galaxies (Eridanus 2, Leo T, Pheonix, NGC 6822, Leo A, and Cetus) that lie in our inferred infall zone between 300 and 840 kpc are clearly not affiliated with M31, considering their large angular separation and distance from M31.

In order to extract the mean radial velocity profile, we model the distribution of radial velocities as a Gaussian distribution with a mean velocity, $\bar v_r(r)$, and a velocity dispersion, $\sigma_r (r)$. To obtain smooth estimates of the two, we adopt an iterative Gaussian process regression~\citep{Rasmussen2005} method, which we briefly outline here but leave further details to Appendix \ref{sec:gp}. Specifically, we first extract a rough estimate of the mean velocity profile $\bar v_r(r)$ assuming a constant $\sigma_r$ using Gaussian process regression. The estimated $\bar v_r(r)$ profile is then combined with the observed velocities to obtain a radial-dependent velocity dispersion profile, $\sigma_r(r)$. Finally, the $\bar v_r(r)$ profile and its uncertainty is refined by fitting a Gaussian process with the estimated $\sigma_r(r)$ profile as the noise term, in addition to a kernel that determines the uncertainty on the mean profile $\bar v_r(r)$. By this process, we self-consistently obtain smooth estimates of $\bar v_r(r)$, $\sigma_r(r)$ as well as the uncertainty on $\bar v_r(r)$.

The fitted $\bar v_r(r)$ profile and its uncertainty are shown in the top panel of \reffig{fig:local_group}.%
\footnote{See also Fig.~11 of \citet{Deason2020} for a similar figure, where the $\bar v_r(r)$ profile was obtained via the Savitzky--Golay smoothing algorithm and a slightly different galaxy sample. However, \citet{Deason2020} focused on the slope of the $\bar v_r$ profile rather than $\bar v_r$ itself.}
The inner part of the profile is flat and consistent with zero net radial flow, as expected for the virialized part of the halo where the density remains largely static. On the largest scale, the positive radial velocity is dominated by the Hubble expansion of the universe. The profile crosses zero at the turnaround radius $\rta \simeq 840\, \kpc$, within which matter starts to fall towards the halo. Within this infall zone but outside the virialized region, the mean $\vr$ profile exhibits a clear minimum that defines the depletion radius $\rd \simeq 560\, \kpc$. The matter in between $\rd$ and $\rta$ is being pumped into the region inside $\rd$, so $\rd$ unveils precisely the border where the MW is feeding on the environment. The amplitude of the maximum infall velocity is relatively small compared to the scatter of the velocities, revealing the MW halo is only growing at a very low rate.

The Gaussian process also enables a probabilistic way to asses the uncertainty in measuring the two characteristics, as it provides a posterior distribution of the entire profile.
We sample $10^4$ random realizations from the posterior of the velocity profile and measure the halo edges respectively.
In most ($>95\%$) realizations, an infall region is detectable with $300\kpc < \rd < 1000 \kpc$.
Taking their average and dispersion, we locate the inner depletion radius at $\rd=559\pm 107\, \kpc$
and turnaround radius at $\rta=839\pm 121\, \kpc$. 
The maximum infall velocity is estimated to be $v_\mathrm{inf, max}=-46_{-39}^{+24}\mathrm{km s^{-1}}$, suggesting that our tentative detection of the infall zone is only at a marginal significance at about 2 $\sigma$ level. 
This is due to both the at most weak infall zone around the MW and the size of the uncertainty given the limited tracer sample size, the latter of which can be reduced by enlarging the nearby galaxy sample in future observation. Despite this, the infall region is also clearly detectable using other smoothing techniques such as the moving average or the Savitzky-–Golay smoothing algorithm \citep{Deason2020}.

It is worth pointing out that the above turnaround radius encloses the M31 (at $r=780 \kpc$),  the MW's massive companion.
Though the M31 and its satellites are excluded from the analysis, 
the M31 can perturb the velocity flow pattern in the vicinity and make the isovelocity surface anisotropic (e.g., \citealt{Deason2020}).
Therefore, our estimate of the turnaround radius should be viewed as a rough estimate in an spherically averaged sense.

\section{Interpreting the measurements with simulations}\label{sec:validate} 

The above measurements are compared with those of simulated halos in the state-of-the-art cosmological hydrodynamical simulation Illustris TNG100 as detailed in Appendix~\ref{sec:simu}.
Following similar procedures to those in \refsec{sec:mw},
for each MW-sized halo in TNG100, 
we identify the turnaround radius, $\rta$, 
as the furthest zero velocity radius along the mean radial velocity profile
and the inner depletion radius, $\rd$, as the furthest local minimum point within $\rta$.

Unlike the MW, for some halos (especially low-mass ones) we fail to locate a $\rd$
beyond the halo virial radius
$\R{vir}$ due to the lack of an infall region in the velocity profile 
(see also e.g., \citealt{Cuesta2008,Fong2020}).
We exclude those halos without a detectable infall zone ($n=1517$) from the parent sample ($n=4681$) of MW-sized halos. We emphasize that the differing strength of the infall zone around halos of a given mass is itself an important diagnostic of the dynamical state and environment of the halo. By definition, halos without an infall region have halted their mass growth while those with one are still accreting.

Our MW is observed to be embedded in a relatively cold environment dynamically, which we find to have a significant influence on the outer halo profile. To make a fair comparison, we select a \emph{fiducial} sample of halos ($n=2153$) with similar masses and dynamical environments to the MW. Out of the fiducial sample, we further select an \emph{LG-like} sample ($n=35$) with the additional requirement of having a close massive companion as detailed in Appendix~\ref{sec:enviro}.

\begin{figure*}[hbtp]
\centering
\includegraphics[width=0.95\textwidth]{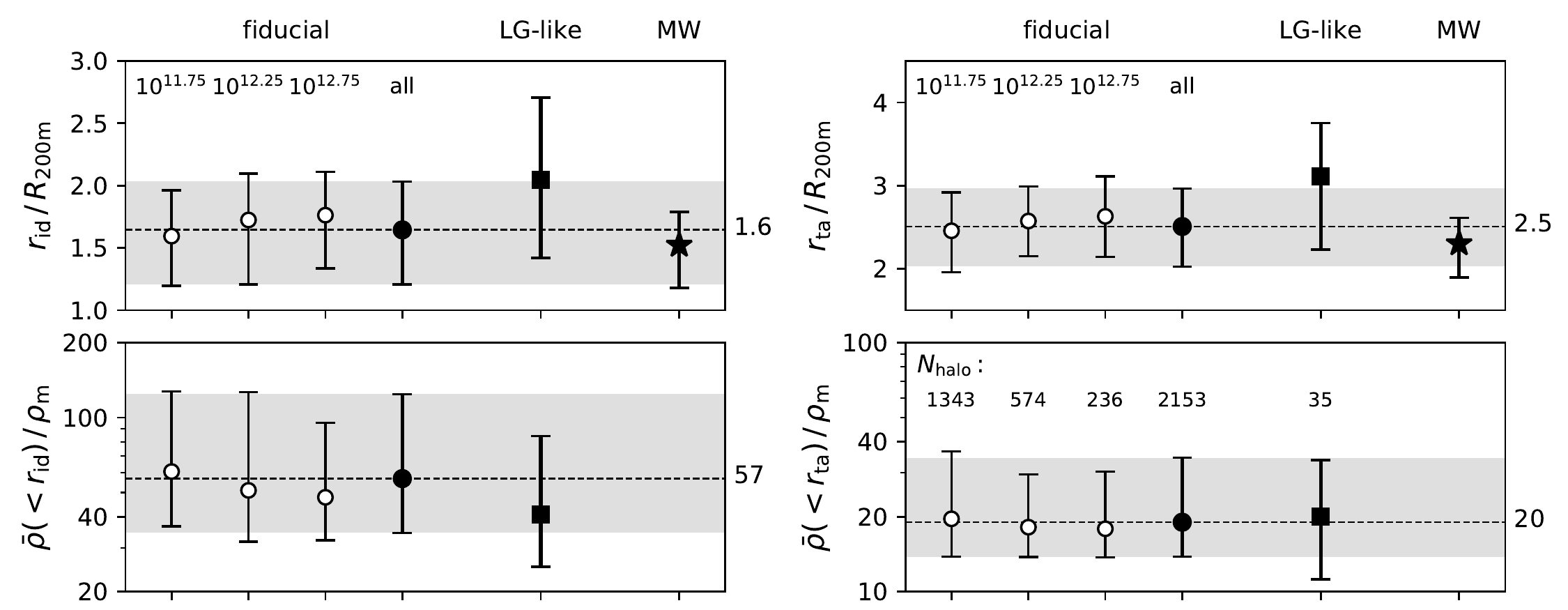}
\caption{%
  Halo edges and corresponding mean enclosed densities of simulated halos.
  The fiducial halo sample and the LG-like (paired) halos
  are shown as solid circles and squares, respectively.
  The fiducial sample is further divided into three halo mass bins, which are shown as open circles.
  The symbols and error bars correspond to the median and the $50\pm34$th percentiles, respectively, with those of the fiducial sample also indicated by horizontal lines and bands for ease of comparison.
  In the top panels, the measurements of MW edges are also shown as star symbols for reference.
}
\label{fig:estimate}
\end{figure*}

\begin{figure}[hbtp]
\centering
\includegraphics[width=0.47\textwidth]{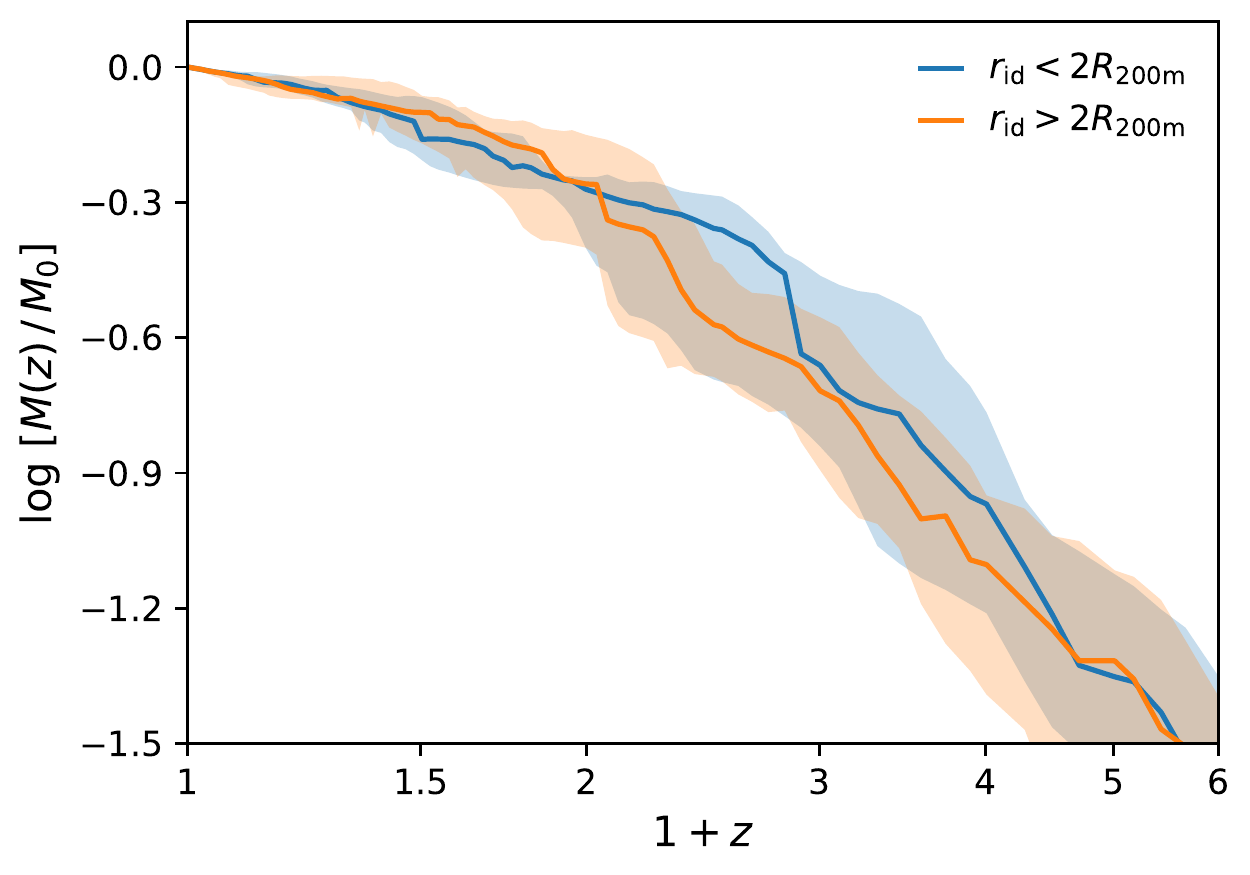}
\caption{%
  Median mass growth history of the LG-like halos.
  The sample is divided into two equal subsets by $r_\mathrm{id}/\R{200m}$.
  The shaded bands show the interval between the 20th to 80th percentiles.
}
\label{fig:mah}
\end{figure}

For the fiducial sample, as shown in \reffig{fig:estimate},
$\rd \sim 1.6 \R{200m}$ with a mean enclosed density $\bar\rho(<\rd) \sim 60 \rhom$
for the whole mass range explored.
It is consistent with \citet{Fong2020}, where $\rd$ is identified using dark matter particles.
As to the turnaround radius, we have $\rta \sim 2.5 \R{200m}$
and $\bar\rho(<\rta) \sim 20 \rhom$ with little mass dependence.
This seems consistent with the idealized spherical collapse model,
which predicts a constant overdensity within $\rta$.
However, \citet{Fong2020} found a much stronger mass dependence
that $\bar\rho(<\rta)$ is significantly higher for smaller halos in their sample.
This is probably because some smaller halos reside in the vicinity of massive halos,
whose tidal force can reduce $\rta$ and thus increase the density enclosed.%
\footnote{%
  Note that $\rd$ is located at a smaller radius, hence less influenced by such effect.
}
In contrast, here we only consider the halos in a dynamically cold environment with a detectable infall zone, resulting in the absence of a mass dependence.

A massive companion halo like the M31 in the LG can perturb the velocity flow pattern in the vicinity and make the isovelocity surface anisotropic (e.g., \citealt{Deason2020}).
Nevertheless, our estimate still provides a meaningful quantification of the halo edges in a spherically averaged sense.
As shown in \reffig{fig:estimate}, the halo edges of LG-like halos are typically slightly larger 
than those in the fiducial sample: $\rd\sim 2\R{200m}$ and $\rta \sim 3\R{200m}$.
It leads to a smaller density within $\rd$, $\bar\rho(<\rd)\sim 40 \rhom$, 
because the density is a decreasing function of radius.
$\bar\rho(<\rta)$ is less affected due to the mass compensated at this scale
by the companion halo.

Based on the characteristic densities enclosed within halo edges,
it is straightforward to estimate the corresponding MW masses,
$\hat M(<\rd) = 60 \rhom \times \frac{4}{3}\pi \rd^3$ and
$\hat M(<\rta) = 20 \rhom \times \frac{4}{3}\pi \rta^3$.
Considering the dispersion among the enclosed densities for the fiducial sample,
the uncertainties in the above mass estimates are about $0.25$ dex and $0.2$ dex, respectively.
Note that this uncertainty actually represents the total uncertainty,
including both the halo-to-halo scatter and the uncertainty due to imprecise determination of the halo edges. The resulting mass estimates are shown in \reffig{fig:local_group}. Calibrating the enclosed densities using the fiducial sample, we obtain $M(<\rd)=1.6_{-0.6}^{+1.9}\times 10^{12} M_\odot$ and $M(<\rta)=1.9_{-0.5}^{+1.5}\times 10^{12} M_\odot$, while calibrating on LG-like halos leads to a lower estimate of $M(<\rd)=1.2_{-0.5}^{+1.2}\times 10^{12} M_\odot$. We emphasize that our new measurements of the outermost edges are independent from the inner halo measurements, and the enclosed masses within these radii can be potentially improved with a better theoretical understanding of the outer halo profile.

The observed $\rd$ and $\rta$ for our MW are about 1.5 and 2.3 times of the $\R{200m,MW}=364\pm19\,\kpc$ measured by \citet{Li2020}.
Surprisingly, this measured $\rd/\R{200m}\approx 1.5$ for our MW is closer to that of the fiducial sample rather than the LG-like halos, even though they are still consistent within the uncertainties. This slight discrepancy may be interpreted as indicating the unique formation history of our MW compared with typical simulated ones in the LG-like sample. To demonstrate this, we divide the LG-like halos into two equal subsamples by $\rd/\R{200m}$ and compare their mass growth histories in \reffig{fig:mah}.
Halos with a smaller $\rd$ seem to have accreted more mass at an earlier time on average, consistent with \cite{Fong2020}.
We have to point out that the limited sample size and the large uncertainties in our current analysis prevent us from making a strong conclusion. Nevertheless, these results illustrate a promising way to constrain the MW formation history in greater details with a larger LG galaxy sample in the future.

\section{Comparisons with previous measurements of the MW mass profile} \label{sec:compare}
In \reffig{fig:local_group}, we also show 
previous measurements of the inner halo profile (see \citealt{Wang2019b} for a more comprehensive review) using halo stars \citep{Xue2008,Huang2016, Ablimit2017, Zhai2018,Deason2020a}, globular clusters 
\citep{Sohn2018,Watkins2018,Vasiliev2018,Eadie2018},
and satellite galaxies \citep{Watkins2010a,Callingham2020,Cautun2019,Li2020,Fritz2020}.
In particular, \citet{Li2020} recently measured the mass profile of the MW outer halo
using satellite galaxies within 300 kpc
and obtained $M_\mathrm{200c}=1.23_{-0.18}^{+0.21}\times 10^{12} M_\odot$ with a concentration of $c=9.4_{ -2.1}^{ +2.8}$.
Assuming an NFW profile, these are equivalent to $M_\mathrm{200m}=1.58\pm 0.25\times 10^{12} M_\odot$ and $\R{200m}=364\pm 19\,\kpc$ or $M_\mathrm{vir}=1.43\pm 0.23\times 10^{12} M_\odot$ and $\R{vir}=297\pm 16\,\kpc$.

Our estimated outer halo masses are further compared with the extrapolated mass profile of the MW based on above measurements of the inner halo mass in Fig.~\ref{fig:local_group}.
It is known that the outer density profiles of halos at $r > \R{200m}$ are remarkably
universal when the radius is normalized by $\R{200m}$ (\citealt{Diemer2014}, see also \reffig{fig:vsig_profile} in Appendix),
which allows us to make profile extrapolation with a reasonable precision.
We rescale the density profiles of the aforementioned simulated halos
to the $\R{200m,MW}$ measured in \citet{Li2020}, 
as $\rho_\mathrm{scaled}(r')=\rho_\mathrm{original}(\frac{r'}{\R{200m,MW}}\R{200m})$. 
To take the uncertainty in $\R{200m,MW}$ into account, the $\R{200m,MW}$ value used to rescale each halo is drawn randomly from the posterior distribution of $\R{200m,MW}$ each time. 
The extrapolated profiles for the fiducial and LG-like halos are quite close within $\rd$,
while the mass at larger scale for the LG-like halos is significantly higher due to the presence of the companion halo.
Both profiles are consistent with the mass estimates at our measured outer edges, although slightly closer to those adopting the fiducial enclosed densities. Note that the enclosed density within $\rd$ depends mostly on the location of $\rd/\R{200m}$ and is not sensitive to how the halo is selected as the profiles are largely universal around this scale. The slightly lower $M(<\rd)$ estimated adopting LG-like depletion density compared with the extrapolated mass profile can thus be viewed as an alternative representation of the smaller $\rd/\R{200m}$ in the observation compared with that of the simulated LG-like halos.

Our measurements of the halo edges are also consistent with previous results at scales beyond the virial radius.
The measured $\rd$ is about 1.9 times of the MW caustic radius $r_\mathrm{sp} = 292 \pm 61 \kpc$ measured by \citet{Deason2020}.
It is consistent with \citet{Fong2020},
who found that $\rd$ is located at $1.7 \sim 2.6$ typically.
The turnaround radius of the MW is slightly smaller than the estimated LG turnaround radius at about $1\mpc$ 
(e.g., \citealt{Karachentsev2002}) as expected.
Our mass estimates within the depletion and turnaround radii are also consistent 
with previous measurements of the ``total'' mass of the Local Group using various methods,
including the timing argument (e.g., \citealt{Li2008},  $5.3_{-1.7}^{+2.5} \times 10^{12} \msun$), simulation-based statistics of the MW-M31 motions (e.g., \citealt{Gonzalez2014}, $ 4.2^{+3.4}_{-2.0}\times 10^{12} \msun$) and dynamical modeling of the Local Group galaxies (e.g., \citealt{Penarrubia2016}, $ 2.64^{+0.42}_{-0.38}\times 10^{12} \msun$;
\citealt{Shaya2017}, $5.15^{\pm0.35} \times 10^{12} \msun$),
though it might be worth further investigating 
what radii such estimates correspond to \citep[e.g.,][]{Penarrubia2017}.

%% =====================================
\section{Summary} \label{sec:conclusion}

Following recent developments in characterizing the boundary of halos \citep{Fong2020},
we measure for the first time the inner depletion radius (aka.\ the maximum infall radius)
and the turnaround radius of the MW halo
and the corresponding enclosed masses. 
This inner depletion radius is expected to be the natural boundary demarcating 
the transition between a growing halo and the environment being depleted,
while the turnaround separates the infall region from the receding background.

Using the radial motion of nearby dwarf galaxies, 
we estimate the inner depletion radius of the MW halo to be $r_\mathrm{id}=559\pm 107 \mathrm{kpc}$ 
and the turnaround radius to be $r_\mathrm{ta}=839\pm 121 \mathrm{kpc}$.
Our detection of the infall zone is only at a marginal significance at about 2 $\sigma$ level due to both the at most weak infall zone around the Milky Way and the size of the uncertainty given the limited tracer sample size. The latter can be improved by enlarging the nearby galaxy sample in future observation. With more observational data, it might also be possible to measure the anisotropic depletion and turnaround surfaces that better characterize the influence from the neighboring M31 halo.

Applying the same analysis to groups with similar masses and dynamical environments in the hydrodynamical simulation Illustris TNG100,  
we find that $\rd \sim 1.6 \R{200m}$ for the whole mass range explored,
while the mean enclosed  density $\bar\rho(<\rd) \sim 60 \rhom$,
which is consistent with \citet{Fong2020}.
As to the turnaround radius, we have $\rta \sim 2.5 \R{200m}$
and $\bar\rho(<\rta) \sim 20 \rhom$ with little mass dependence.
The constant enclose density allows us to estimate the MW mass as
$M(<\rd)=1.6_{-0.6}^{+1.9}\times 10^{12} M_\odot$ and $M(<\rta)=1.9_{-0.5}^{+1.5}\times 10^{12} M_\odot$. The estimates are consistent with the mass profile constrained 
at smaller radii \citep[e.g.,][]{Li2020} and the LG mass constrained at larger scale (see \refsec{sec:compare}).

When the selection criterion in the simulation is tightened to select paired halos similar to our LG, we find a slightly lower enclosed density associated with a larger maximum infall radius. Taking this enclosed density as a reference,
the MW mass within the boundary is expected to be 
$M(<\rd)=1.2_{-0.5}^{+1.2}\times 10^{12} M_\odot$.

These measurements directly quantify the ongoing evolution of the MW outer halo and provide constraints on the current dynamical state and mass content of the MW independent of internal dynamics. The tentative detection of the infall zone indicates the MW halo is likely still accreting mass from the surrounding environment rather than having halted its growth. The slightly smaller $\rd/\R{200m}$ of the observed MW halo compared with those of LG-analogies in the simulation can be potentially interpreted as suggesting the distinct formation history of our MW. For example, a smaller $\rd/\R{200m}$ indicates a smaller infall region and might imply inefficient recent mass growth. However, we emphasize that the difference is still very weak given the large measurement uncertainties, which could be improved with a larger nearby galaxy sample in the future. Improved theoretical understandings of the connections between $\rd$ and other halo properties can also help to better interpret the measurements. Nevertheless, we expect these measured outermost characteristic radii can serve as alternative selection variables when placing our MW halo into the context of cosmological galaxy formation.

\acknowledgments

% \vspace*{1cm}
We are very grateful to Hongyu Gao, Matthew Fong and Wenting Wang for their helpful discussions.

This work is supported by NSFC (11973032, 11890691, 12022307, 11621303), National Key Basic Research and Development Program of 
China (No.\ 2018YFA0404504), 111 project (No.\ B20019), and the science research grants from the China Manned Space Project (No.\ CMS-CSST-2021-A03, CMS-CSST-2021-B03). We gratefully acknowledge the support of the MOE Key Lab for Particle Physics,
Astrophysics and Cosmology, Ministry of Education. The computation of this work is partly done on the \textsc{Gravity} supercomputer at the Department of Astronomy, Shanghai Jiao Tong University. 

The IllustrisTNG simulations were undertaken with compute time awarded by the Gauss Centre for Supercomputing (GCS) under GCS Large-Scale Projects GCS-ILLU and GCS-DWAR on the GCS share of the supercomputer Hazel Hen at the High Performance Computing Center Stuttgart (HLRS), as well as on the machines of the Max Planck Computing and Data Facility (MPCDF) in Garching, Germany.

This research has made use of NASA's Astrophysics Data System
and adstex (\url{https://github.com/yymao/adstex}).

\textit{Software:} 
  Astropy \citep{AstropyCollaboration2013},
  scikit-learn \citep{Pedregosa2011},
  Numpy \citep{Walt2011}, 
  Scipy \citep{Oliphant2007},
  Matplotlib \citep{Hunter2007}

% \balance

%% =====================================

\appendix
\counterwithin{figure}{section}
\section{The mean and dispersion profile of radial velocity} \label{sec:gp}

We calculate the mean radial velocity profile, $\bar v_r(r)$, from the nearby galaxies,
using the Gaussian process regression \citep{Rasmussen2005}
implemented in the public code \textsc{scikit-learn} \citep{Pedregosa2011}.
The Gaussian process is particularly suitable for this problem.
In the view of Gaussian process,
the $\{r, v_r\}$ sample of subhalos can be naturally seen as the outcome
of an underlying mean function, $\bar v_r(r)$,
convolved with a random scatter, $\sigma_r(r)$ 
(e.g., due to the internal motion of the halos and filaments).
This scatter is known as the white noise term in the framework of Gaussian process.

Considering that the distance, $r$, can span several orders of magnitude, 
we perform the regression as a function of $x=\log_{10}(r+r_0)$,
where $r_0$ is added to avoid the possible divergence or artificial fluctuation at very small $r$.
We adopt $r_0=40\kpc$ for the Milky Way and $0.1\R{200m}$ for simulated halos.
Nevertheless, we find that the inferred maximum infall radius is not sensitive to the choice of $r_0$ at all.
A Gaussian process is completely characterized by the prior mean function, $m(x)$, 
and the kernel function, $k(x, x')$,
where $k$ determines the correlation among the functional values at arbitrary positions.
We take the zero mean function as prior 
and adopt the following kernel,
\begin{equation}
  k(x, x') = k_0 S(x-x') + \sigma_r^2(x)\delta(x-x'),
\end{equation}
which combines a Gaussian kernel, $S$, with a tunable amplitude, $k_0$,
and a white noise, $\sigma_r^2$.
The Gaussian kernel is controlled by the scale length $s$,
\begin{equation}
  S(x) = \exp\big(- 0.5{x^2}/{s^2} \big). \label{eq:SE}
\end{equation}
We fix the parameters $s=0.2$, 
which is found to be optimal for the mean stacked $\vr$ profile of the fiducial halo sample.
Allowing it to be a free parameter during fitting individual halos generally leads to noisier result due to the limited sample size.
Nevertheless,
the result is robust against moderate changes of $s$.
Varying $s$ from 0.15 to 0.3 only changes the median value of $\bar\rho(<\rd)$ or $\bar\rho(<\rta)$
of the halos by a factor of $\lesssim 10\%$, which is much smaller than the halo-to-halo scatter.
Using a significantly smaller or larger $s$ may lead to apparent under-smoothing or over-smoothing.

For each dataset, once the best-fit $k_0$ is determined by maximizing the likelihood of the input data,
we can predict $\bar v_r$ and its uncertainty $\sigma_{\bar v_r}$%
\footnote{Note that $\sigma_{\bar v_r}$ represents the uncertainty of $\bar v_r$,
which is different from the actual velocity dispersion, $\sigma_{r}$.}
and covariance matrix at arbitrary position and measure the halo edges.
Moreover, Gaussian process also allows one to sample random realizations of the velocity profiles
around the mean function, which is convenient for taking the statistical uncertainty due to limited sample size into account.
We refer the interested readers to \citet{Rasmussen2005} for details.

Note that the velocity dispersion profile, $\sigma_{r}(r)$, appears in the Gaussian process
as the noise term. 
It can be directly measured from data but only when the mean velocity profile $\bar v_r(r)$ is determined first. We resolve it with a two-step iteration. 
We first assume a constant dispersion by forcing $\sigma_r^2= k_0$ during the fitting and predict $\bar v_r(r)$. 
A more realistic $\sigma_{r}$ is then calculated from data through kernel smoothing
and used for retraining the Gaussian process.
For simplicity, here we use the same smoothing kernel in \refeqn{eq:SE},
\begin{equation}
  \sigma_r^2(x) = \frac{\textstyle \sum_i S(x-x_i) \left[v_{r,i} - \bar v_r(x)\right]^2}{\textstyle\sum_i S(x-x_i)}.
\end{equation}
Unlike the binned statistics,
this approach can output a smooth $\sigma_r^2(r)$ profile even when the number of tracers is small.

%% =====================================
\section{Simulation data} \label{sec:simu}

We use the hydrodynamical simulation Illustris TNG100 to study the edges of Galactic-sized halos.
The TNG100 is one of the IllustrisTNG cosmological simulation suite 
\citep{Nelson2019,Springel2018,Pillepich2018,Nelson2018,Naiman2018,Marinacci2018}
performed with comprehensive prescriptions for various physical processes in galaxy formation.
The TNG100 was carried in a periodic box of size $L=75\mpch$ with
the Planck Collaboration XIII \citep{PlanckCollaboration2016} cosmology,
$\Omega_\mathrm{m} =  0.3089$, $\Omega_\mathrm{b} = 0.0486$, $\Omega_{\Lambda} = 0.6911$, 
$H_0= 67.74\, \kms \mpc^{-1}$, $n_\mathrm{s}=0.9667$, and $\sigma_8 = 0.8159$.
The baryonic mass resolution is $9.44\times10^5\msunh$ and the dark matter particle mass is $5.06\times10^6\msunh$.

\citet[see also \citealt{More2015,Shi2016a}]{Diemer2014}
argued that $R_\mathrm{200m}$ is a good choice to scale the halo structure at large radii to obtain unified profiles.
In the following analysis, 
we will use the quantities defined at $\R{200m}$ (e.g., the enclosed mass $\M{200m}$ and the circular velocity $\V{200m}=\sqrt{G\M{200m}/\R{200m}}$) as references when studying the halo profiles.

We select a parent sample ($n=4681$) of halos with $\M{200m} \in [10^{11.5}, 10^{13}] \msun$
along with their nearby subhalos within $8\R{200m}$.
The choice of $8\R{200m}$ is guided by
the observation data used in this work ($3\mpc\! \sim\! 8\R{200m, MW}$).
To ensure a similar statistical power to the observations,
all subhalos within this radius are included regardless of their luminosity.
This is not a concern because the kinematics of subhalos or dwarf galaxies 
are expected to be insensitive to their masses or luminosities at the interested scales.
We have confirmed that using luminous subhalos with at least one star particle
indeed produces consistent results but with mildly larger statistical uncertainties.

To mimic the treatment of the LG observation
and alleviate the contamination from nearby massive halos, 
for each target halo
we further identify its largest neighboring halo within $8\R{200m}$
and remove the subhalos within $1.5\R{200m,ngb}$ of this neighboring halo.
Tests show that such a treatment can give more robust estimates of the edges
when the neighboring halo is close.
Moreover, as mentioned in \refsec{sec:validate},
we exclude the halos without a detectable infall zone ($n=1517$) from the parent sample.

%% =====================================
\subsection{Selecting MW analogies} \label{sec:enviro}

We are particularly interested in halos similar to our MW and the LG system.
The LG is a quite isolated group where the MW and M31 are the largest two halos within $3 \mpc$ 
\citep[e.g.,][]{Shaya2017}.
In many studies (e.g., \citealt{Wang2017b}),
people characterize the level of isolation by the distance to more massive halos.
However, we notice that even without massive halos,
the environment sometimes can still be dynamically hot due to the filaments,
which can accelerate galaxies and result in a large velocity dispersion.
For this reason, we introduce a local environment parameter, $\env$,
and use it to select MW halo analogies in the simulation.

We define $\env$ for each halo as the ratio between
the velocity dispersion outside and inside the halo.
Specifically, the velocity dispersion profile,
$\sigma_r(r)$, around each halo is first computed through kernel smoothing as detailed in the Appendix~\ref{sec:gp}.
Taking $r>\rd$ as the environment and $r<0.5\rd$ as the halo region, the $\env$ parameter is defined as\footnote{%
  Note that here $\env$ actually represents the strength of the large-scale structure within $8\R{200m}$ excluding the largest neighbor, because the member galaxies of the neighbor have been removed.
  Nevertheless, $\env$ can still correlate with the size of the neighbor, which is often associated with large-scale structures such as filaments.
}
\begin{equation}
\env = \max\{\sigma_{r} \,|\, r > \rd\} / \max\{\sigma_{r} \,|\, r < 0.5\rd\}.    
\end{equation}
We also checked that using $2\R{200m}$ instead of $\rd$ gives nearly identical values.

As a reference, the nearby dwarf galaxies
have $\max\{\sigma_r\} \simeq 130 \kms$ within $300\kpc$ of the MW 
and $\sim 75 \kms$ beyond $600 \kpc$,
resulting in $\env \sim 0.6$ for the MW. Accordingly, we select a sample of halos in dynamically cold environments with $\env<1$ as the fiducial sample ($n=2153$) for comparison. 
Among them, we further select a subsample of paired LG-like halos ($n=35$)
whose largest neighbor within $8\R{200m}$ has a
mass $0.7<\M{200m,ngb}/\M{200m}<2$ and a distance $1.5<d_\mathrm{ngb}/\R{200m}<3$,
where the subscript `ngb' denotes the neighbor. 

\begin{figure}[bt!]
\centering
\includegraphics[width=0.47\textwidth]{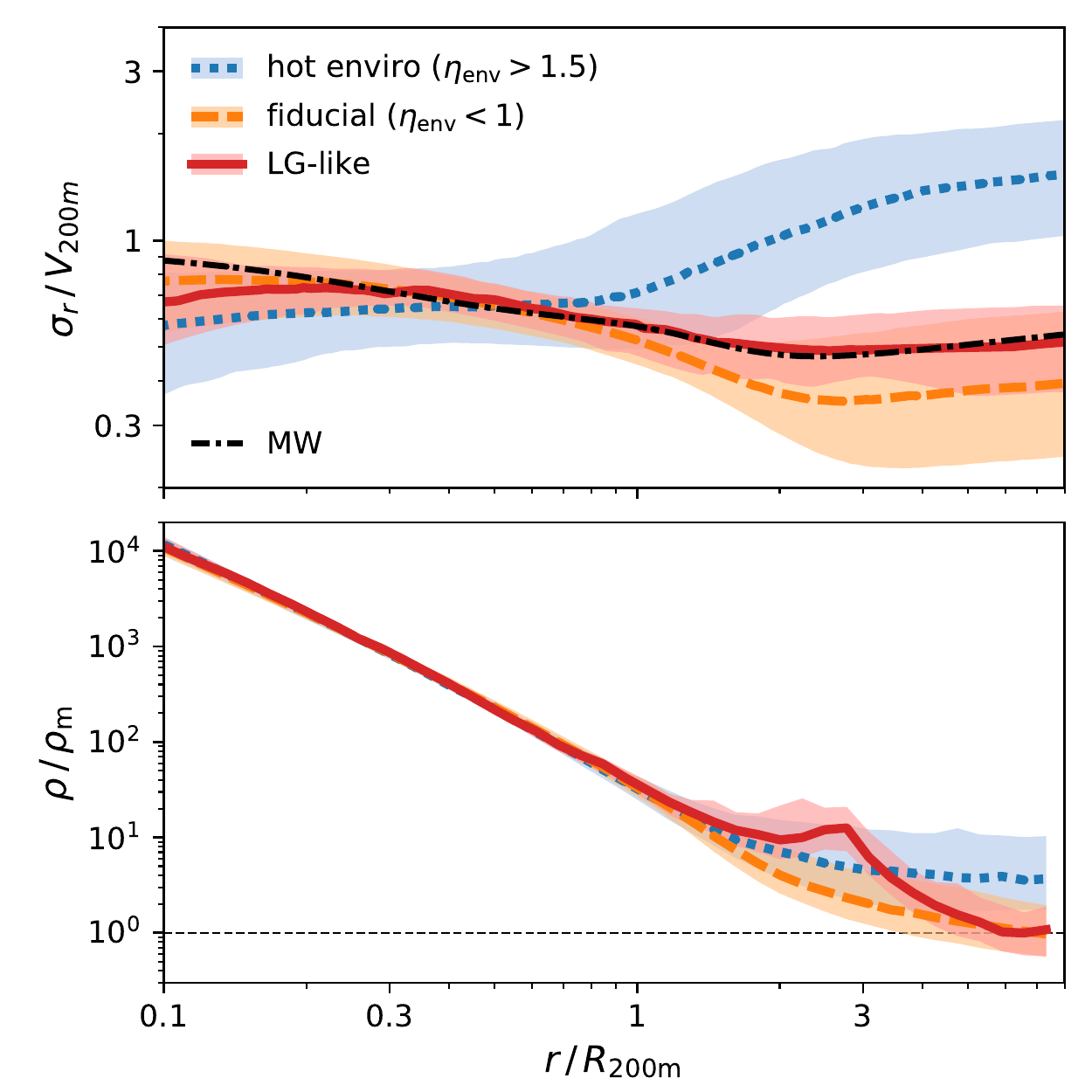}
\caption{%
  Illustration of the environment parameter:
  the median and 68\% quantiles of the velocity dispersion profile (top panel) 
  and density profile (bottom panel) for halos within different environments.
  Two halo samples are selected by the environment parameter ${\env}$
  and labeled as ``hot enviro'' (${\env}>1.5$) and ``fiducial'' (${\env}<1$).
  A subset of the fiducial halos akin to the MW are further chosen 
  by the presence of massive close companion (see text for detail) and labeled as ``LG-like''.
  The velocity dispersion profile of the MW is shown as dash-dotted line for comparison.
}
\label{fig:vsig_profile}
\end{figure}

In Fig.~\ref{fig:vsig_profile},
we show the median radial velocity dispersion profiles and the density profiles
of the above halo samples.
A halo sample in dynamically hot environment with $\env>1.5$ is also shown for comparison.
Remarkably, the dispersion profile of the MW (scaled by $\R{200m}$ measured by \citealt{Li2020})
nearly coincides with that of the LG-like sample, especially for $r>0.2\R{200m}$.
As demonstrated in the bottom panel,
while the median density profiles of different samples are almost identical in the inner halo,
they differ significantly on scales beyond the virial radius.
The profile of the LG-like sample differs from that of the fiducial one at $2\sim 4 \R{200m}$ due to the presence of the massive companion halo. The two coincide with each other again on larger scale for their similar $\env$ values,
lying below the sample with higher $\env$.
Clearly, $\env$ is indeed an indicator of the mass density on large scale,
even for the paired LG-like halos.

%% =====================================
% \newpage
\bibliography{mw_radius}

\end{document}